\documentclass[aps,preprint,superscriptaddress,amsmath,amssymb]{revtex4}

\usepackage{graphicx}
\usepackage{dcolumn}
\usepackage{bm}

\begin{document}

\title{Enhanced photon-extraction efficiency from deterministic quantum-dot microlenses}

\author{M.~Gschrey}
\affiliation{Institut f\"{u}r Festk\"{o}rperphysik, Technische Universit{\"a}t Berlin, Hardenbergstra{\ss}e 36, D-10623 Berlin, Germany}
\author{M.~Seifried}
\affiliation{Institut f\"{u}r Festk\"{o}rperphysik, Technische Universit{\"a}t Berlin, Hardenbergstra{\ss}e 36, D-10623 Berlin, Germany}
\author{L.~Kr\"uger}
\affiliation{Institut f\"{u}r Festk\"{o}rperphysik, Technische Universit{\"a}t Berlin, Hardenbergstra{\ss}e 36, D-10623 Berlin, Germany}
\author{R.~Schmidt}
\affiliation{Institut f\"{u}r Festk\"{o}rperphysik, Technische Universit{\"a}t Berlin, Hardenbergstra{\ss}e 36, D-10623 Berlin, Germany}
\author{J.-H.~Schulze}
\affiliation{Institut f\"{u}r Festk\"{o}rperphysik, Technische Universit{\"a}t Berlin, Hardenbergstra{\ss}e 36, D-10623 Berlin, Germany}
\author{T.~Heindel}
\affiliation{Institut f\"{u}r Festk\"{o}rperphysik, Technische Universit{\"a}t Berlin, Hardenbergstra{\ss}e 36, D-10623 Berlin, Germany}
\author{S.~Burger}
\affiliation{Zuse-Institut Berlin (ZIB), Takustrasse D-714195 Berlin, Germany}
\author{S.~Rodt}
\affiliation{Institut f\"{u}r Festk\"{o}rperphysik, Technische Universit{\"a}t Berlin, Hardenbergstra{\ss}e 36, D-10623 Berlin, Germany}
\author{F.~Schmidt}
\affiliation{Zuse-Institut Berlin (ZIB), Takustrasse D-714195 Berlin, Germany}
\author{A.~Strittmatter}
\affiliation{Institut f\"{u}r Festk\"{o}rperphysik, Technische Universit{\"a}t Berlin, Hardenbergstra{\ss}e 36, D-10623 Berlin, Germany}
\author{S.~Reitzenstein}
\email{stephan.reitzenstein@physik.tu-berlin.de}
\affiliation{Institut f\"{u}r Festk\"{o}rperphysik, Technische Universit{\"a}t Berlin, Hardenbergstra{\ss}e 36, D-10623 Berlin, Germany}

\date{\today}

\maketitle

{\bf The prospect of realizing building blocks for long-distance quantum communication is a major driving force for the development of advanced nanophotonic
devices~\cite{Gis07}. Significant progress has been achieved in this field with respect to the fabrication of efficient quantum-dot-based single-photon 
sources~\cite{Mic00, Pel02, Yua02, Cla10, Hei10, Rei11}. More recently, even spin-photon entanglement~\cite{Gre12, Gao12} and quantum teleportation~\cite{Gao13, Ste13}  have been demonstrated in semiconductor systems. These results are considered as crucial steps towards the realization of a quantum repeater. The related work has almost exclusively been performed on self-assembled quantum dots (QDs) and random device technology. At this point it is clear that further progress in this field towards real applications will rely crucially on deterministic device technologies which will, for instance, enable the processing of bright quantum light sources with pre-defined emission energy~\cite{Dou10}.  

Here we report on enhanced photon-extraction efficiency from monolithically integrated microlenses which are coupled deterministically to single QDs. 
The microlenses with diameters down to 800 nm were aligned to single QDs by in-situ electron-beam lithography using a low-temperature cathodoluminescence setup. 
This deterministic device technology allowed us to obtain an enhancement of photon extraction efficiency for QDs integrated into microlenses as compared to QDs in unstructured 
surfaces. The excellent optical quality of the structures is demonstrated by cathodoluminescence and micro-photoluminescence 
spectroscopy. A Hong-Ou-Mandel experiment states the emission of single indistinguishable photons.
}

An important figure of merit of all non-classical photon sources is the photon-extraction efficiency (PEE), i.e. the probability of coupling a photon emitted by a 
QD into the external optics~\cite{Wak02}. In the case of InGaAs QDs embedded in a GaAs matrix, this measure is limited by total internal reflection to about 1\% for 
collecting optics with a numerical aperture (NA) of 0.8. While the PEE can be enhanced also by advanced waveguide structures or microcavities, the 
microlense-approach provides a simple and, therefore, very appealing concept to increase the photon flux of QD based quantum devices~\cite{Bar02}. Moreover, microlenses 
provide broadband enhancement of the PEE without the need of complicated spectral tuning methods, and as such they are particular attractive for boosting the 
extraction of polarization-entangled photon pairs from the biexciton-exciton cascade of QDs with diminishing fine-structure splitting (FSS)~\cite{Benson2000}. The microlense 
approach is very general and can also be applied in a straightforward way to enhance the PEE of materials such as wide-bandgap II/VI-semiconductors or InGaAs QDs 
on (111)-oriented GaAs with FSS close to zero~\cite{Sch09,Sin09, Jus13} for which the growth of cavity structures is still challenging or has not been mastered yet 
at all. More than this, if combined with suitable cavity structures, significant further enhancement of PEE is expected for the microlens approach~\cite{Bar02}.
 
The application of solid immersion lenses (SILs) to enhance the PEE of photons emitted by a single QD in a semiconductor matrix was nicely discussed in Ref.~\cite{Bar02}. Basically SILs circumvent restrictions imposed by total internal reflection at the semiconductor-air interface which lead to poor PEEs. Using commercially available mm-scale hemispheric SILs with a refractive index in between those of the semiconductor material and air the PEE is enhanced to 6 \%. Significantly better performance and PEE = 17 \% for a NA of the collecting optics of 0.8 can be achieved by integrating the lens directly into the semiconductor material. However, this route is technologically more demanding since it is not trivial to find suitable processing parameters to ''shape'' the integrated SILs and, moreover, to align them to a single target emitter in an ensemble of self-assembled QDs.
  
In this work we apply a recently developed in-situ electron-beam lithography technique \cite{Gsc13} to take advantage of the SIL concept to enhance the PEE. This lithography 
technique which is based on low-temperature cathodoluminescence spectroscopy allows us to realize deterministic and monolithically integrated QD microlenses which are
spatially aligned precisely to single target QDs. While our integrated microlense approach can in principle be adopted to all semiconductor materials with embedded nanostructures, we have chosen the technically mature InGaAs/GaAs material system for demonstration purposes. We have also chosen a generic sample layout without additional optical elements such as a distributed Bragg reflector to focus solely on the geometrically enhanced PEE obtained by deterministically integrated microlenses.

Lens fabrication is sketched in Fig.~\ref{fig:figure1}(a-d). In the central processing step we invert the electron-beam sensitive resist polymethyl methacrylate (PMMA) covering the semiconductor material selectively at the positions of target QDs (see Supplementary Information for details on the sample growth). After development, the locally inverted resist acts as an etch mask which is transferred into the semiconductor material in the subsequent plasma etching step. Here, the exposure dose profile determines the local thickness of the PMMA-etch mask which allows to tailor the microlens shape. A respective calibration curve obtained under an acceleration voltage of 15~kV is presented in Fig.~\ref{fig:figure2} (black curve) and shows that the thickness of the remaining PMMA increases from 0~nm at 10.5~mC/cm$^2$ to 70~nm at 25.5~mC/cm$^2$. This thickness determines the corresponding etch depth which varies by almost 500 nm as a function of the dose as can be seen by the red trace in the same figure. Thus, in order to shape microlenses into the semiconductor material we use the calibration curve presented in Fig.~\ref{fig:figure2} to calculate radial profiles of the exposure dose as indicated in Fig.~\ref{fig:figure1}(b): the dose equals to 26 mC/cm$^2$ in the center of the lens and gets reduced towards the edge following a Gaussian dose profile in radial direction. The resulting height profile of a processed lens as measured by atomic force microscopy is presented as inset in Fig.~\ref{fig:figure2}. Fitting this profile gives an almost ideal Gaussian shape. This result clearly states that our electron-beam lithography approach allows us to write nanostructures under a very precise variation of the exposure dose into the resist - a task which is certainly not feasible with optical lithography which has significantly lower lateral resolution and for which the profile of the dose is essentially fixed and pre-determined by the profile of the laser beam. 

The successful fabrication of our QD-microlenses and the resulting enhancement of the PEE is verified by high-resolution CL-spectroscopy maps. A representative CL-intensity map after processing of the nanostructures is depicted in Fig.~\ref{fig:figure3}. It shows two bright spots on the right hand side where microlenses were fabricated aligned to two selected QDs and many weak spots in the left hand side of the figure associated with emission of QDs in the unprocessed region of the sample. A thorough analysis yields an enhancement factor as high as 6.6 $\pm$ 2.7 (see Supplementary Information).

The theoretical limit for the enhancement factor of a given setup is derived from simulations of Maxwell's equations (see Supplemntary Information). Figure~\ref{fig:figure4}(a) shows the photon extraction efficiency for a plain surface and for a Gaussian lens as a function of the NA of the light-collection optics (the inset gives the light intensity distribution around the lens). The respective enhancement factor is displayed in panel~(b). For a NA of 0.8 which corresponds to the CL setup we obtain an enhancement factor of 9.6. This value is in good agreement with the above given experimental factor of 6.6 $\pm$ 2.7.

Finally, we demonstrate the high optical quality of the QD-microlenses and the potential to act as non-classical light sources in the field of quantum communication. For this purpose we performed high-resolution $\mu$PL studies under CW excitation in order to address the single photon purity of emission via Hanbury-Brown and Twiss (HBT) measurements and the degree of photon indistinguishability via Hong-Ou-Mandel (HOM) two-photon interference experiments (see Supplementary Information). The results are presented in Fig.~\ref{fig:figure5}. The $\mu$PL spectrum of a single QD-microlens is depicted in Fig.~\ref{fig:figure5}(a) and shows discrete emission lines which are identified by photon cross-correlation measurements (not shown here) to stem from the recombination of negatively charged excitons $X^{-}$, biexcitons $XX$, and neutral excitons $X$. Figure~\ref{fig:figure5}(b) shows the photon auto-correlation function $g^{(2)}(\tau)$ of the $X^{-}$ line. The value of $g^{(2)}(\tau)$ at zero delay yields single-photon emission with a very high suppression of multi-photon emission events associated with a measured value of 0.19.  Taking the timing resolution of the setup into account we extract a deconvoluted value of $g^{(2)}(0)<0.001$. Furthermore, the radiative lifetime $\tau_{R}=779$~ps is extracted from the fit.

A high degree of indistinguishability of photons emitted from the QD-microlenses is proven by analyzing the emission of the $X^{-}$ line via a fiber-coupled HOM setup. We performed measurements both for parallel and orthogonal configuration of the half-wave plate, not-switching or switching the polarization in one interferometer arm with respect to the other. For the parallel configuration, presented in Fig.~\ref{fig:figure5}(b), we observe an antibunching dip at zero delay clearly below 0.5 indicating indistinguishability of the emitted photons. The solid line shows a fit to the experimental data according to \cite{Patel2008} and considering the timing resolution of the setup. We determine the coherence time $\tau_{C}=749$~ps, $g^{(2)}_{ \parallel}(0)=0.28$ and a visibility of two-photon interference as high as $V(0)= 0.45$. The visibility is a function of the overlap of the wavefunctions and quantifies the indistinguishability of consecutively emitted photons. For a perfect single-photon source $V$ can be written as $V(\tau)=\frac{ g^{(2)}_{ \bot}(\tau)- g^{(2)}_{ \parallel}(\tau)}{ g^{(2)}_{ \bot}(\tau)}$, being 1 for an ideal emitter. The visibility $V$ is mainly limited by the non-resonant excitation and we expect significantly higher values under quasi- and strict-resonant excitation of the QD~\cite{Legero2003, Ates2009a, He2013, Joens2013}. In Fig.~\ref{fig:figure5}(c) the result of the orthogonal configuration is also presented for reference, showing $g^{(2)}_{ \bot}(0)=0.5$ which confirms the perfect single photon emission of the QD.

In summary, we have demonstrated a versatile method to enhance the photon-extraction efficiency for semiconductor nanostructures by fabricating deterministic microlenses. The QD-microlens structures have proven that our approach has a high potential to boost the application of quantum light sources. Due to their broadband enhancement of photon extraction efficiency and the straightforward fabrication process they will be particularly appealing for the realization of polarization entangled photon pairs and sources of indistinguishable photons in quantum repeater networks.   

\subsection*{Methods}

The deterministic microlenses are fabricated in the following way. First, we spin-coat a 190~nm thick layer of PMMA on the sample before recording cathodoluminesence intensity maps in the CL-system at low temperature and a low dose of 10.5 mC/cm$^2$. Out of a large number of embedded QDs we select target QDs which show reasonably spatially isolated emission spots, and which are therefore suitable to be deterministically integrated into microlenses. In the subsequent in-situ electron-beam lithography step we write lens-patterns into the resist as described above. Actually, we are writing concentric circles which are centered at the position of a target QD under variation of the dose (highest dose at the center and lower doses towards the edge of the lens) to form the correct dose profile as sketched in Fig.~\ref{fig:figure1}. We would like to point out that the whole selection and in-situ lithography process of typically 10 deterministic QD-microlenses in a write-field of 23.9 to 17.9 $\mu$m is performed within 4 minutes without moving the sample. Afterwards, the sample is transferred out of the CL-system and dry etching is performed in a ICP-RIE plasma under a pressure of 0.08 Pa with 100 W ICP coil power and -215 V substrate bias voltage. A combination of Cl$_2$:BCl$_3$:Ar, with a ratio of 1:3:1, is used to reach a selectivity of GaAs against unexposed PMMA of 2. Realizing an etching depth of about 550 nm, the lens profile was transfered from the inverted PMMA into the semiconductor and the QD layer around the lenses was removed.

The lens shape was deliberately chosen to be of Gaussian type: Three-dimensional electron beam lithography is quite challenging as he spatial dose profile must resemble the targeted lens profile. This is complicated by the omnipresent proximity effect in EBL whose local impact can be described by a Gaussian profile~\cite{rishton87}. As the convolution of a Gaussian (lens) profile with another Gauss is still Gaussian, the targeted shape will be maintained in this way. 

\subsection*{Acknowledgements}
We acknowledge support from Deutsche Forschungsgemeinschaft (DFG) through SFB 787 ''Semiconductor Nanophotonics: Materials, Models, Devices'' (project C12) and from German-Israeli-Foundation for Scientific Research and Development, Grant-No.: 1148-77.14/2011.

\subsection*{Author contributions}
M.G. and M.S. performed the CL experiments and the CL lithography, L.K. and T.H. carried out the $\mu$PL/HOM investigations. M.S. conducted the numerical modelling with support from S.B and F.S.. R.S. processed the samples in the clean room. J.-H.S. and A.S. grew the samples. S.R. and S.R. initiated the research, supervised the project and wrote the manuscript with input from all authors. 

\clearpage

\bibliography{Gschrey}
%\bibliographystyle{apsper}

%\end{thebibliography

\clearpage

\noindent Figure Captions

\begin{figure}[h]
\centering
\caption{(Color online) Schematic view of the lens fabrication process. (a) The sample's luminescence is mapped by cathodoluminescence spectroscopy. Along this, the resist is exposed to an electron dose of up to 11~mC$\cdot$cm$^{-2}$ and gets soluble upon development. (b)  On top of suitable QDs, lens structures are written into the resist by interlinking the afore cracked PMMA chains by using an additional electron dose. (c) Singly exposed resist is removed by applying the solvent methylisobutylketon (MIBK) and the lens shape forms in the inverted regions. (d) Upon dry etching the lens profile is transferred from the inverted PMMA into the semiconductor.}
\label{fig:figure1}
\end{figure}

\begin{figure}[h]
\centering
\caption{(Color online) Prerequisite for etching lens structures is to create 3D lens profiles in the resist. Left axis, black curve: Height profile of 190 nm PMMA after development as a function of pre-applied electron dose. Between 0 and 20~mC$\cdot$cm$^{-2}$ we observe the characteristics of a positive tone resist. Then it starts to reinterlink and the remaining height is a function of the applied dose. Right axis, red curve: Etch depth in GaAs that corresponds directly to the resist thickness as given by the black curve. The resist profile is precisely transferred into the semiconductor as proven by AFM measurements of etched lenses (inset) that resemble the targeted Gaussian profiles.}
\label{fig:figure2}
\end{figure}

\begin{figure}[h]
\centering
\caption{(Color online) CL luminescence map of a sample that contains two lenses (right half-plane) in direct vicinity to an unprocessed area (left half-plane). The scheme on top illustrates the (exaggerated) height profile along the dashed white line. The photon-extraction enhancement due to the lenses is directly evident by comparing both half-planes.}
\label{fig:figure3}
\end{figure}

\begin{figure}[h]
\centering
\caption{(Color online) (a) Theoretical results for the photon-extraction efficiency from a plain surface (black line, dots) and from a Gaussian lens structure (red line, diamonds) as a function of the NA of the light-collection optics. The inset displays the light-intensity distribution around the lens. (b) Enhancement factor for a Gaussian lens.}
\label{fig:figure4}
\end{figure}

\begin{figure}[h]
\centering
\caption{(Color online) $\mu$PL and HOM characterization of a microlens-boosted QD. (a) Emission lines from the single QD that stem from the negatively-charged exciton (X$^-$), the biexciton (XX) and from the neutral exciton (X). Inset: $g^{2}(\tau)$ results for the X$^-$ emission yielding a $g^{2}(0) < 0.001$ by fitting the experimental data. (b) and (c) HOM measurement for parallel polarization configuration and for perpendicular configuration, respectively. Scattered symbols represent experimental data and blue lines are fitted curves as described in the text.}
\label{fig:figure5}
\end{figure}

\clearpage

\begin{figure}[h]
\centering\includegraphics[width=8 cm]{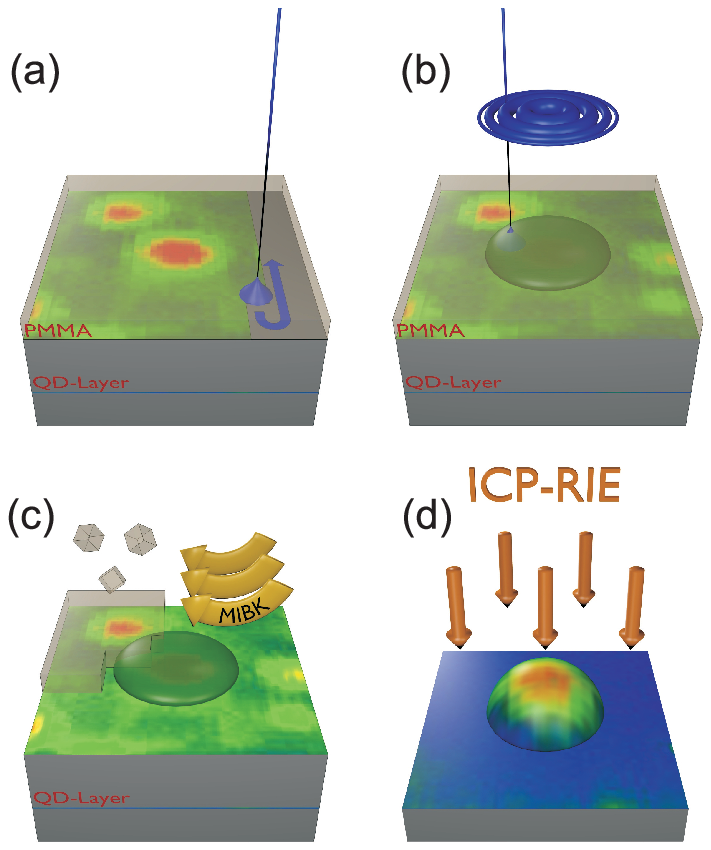}
\end{figure}
\begin{center}Gschrey et al. Fig. 1.\end{center}

\clearpage

\begin{figure}[h]
\label{fig:cal_curve}
\centering\includegraphics[width=8 cm]{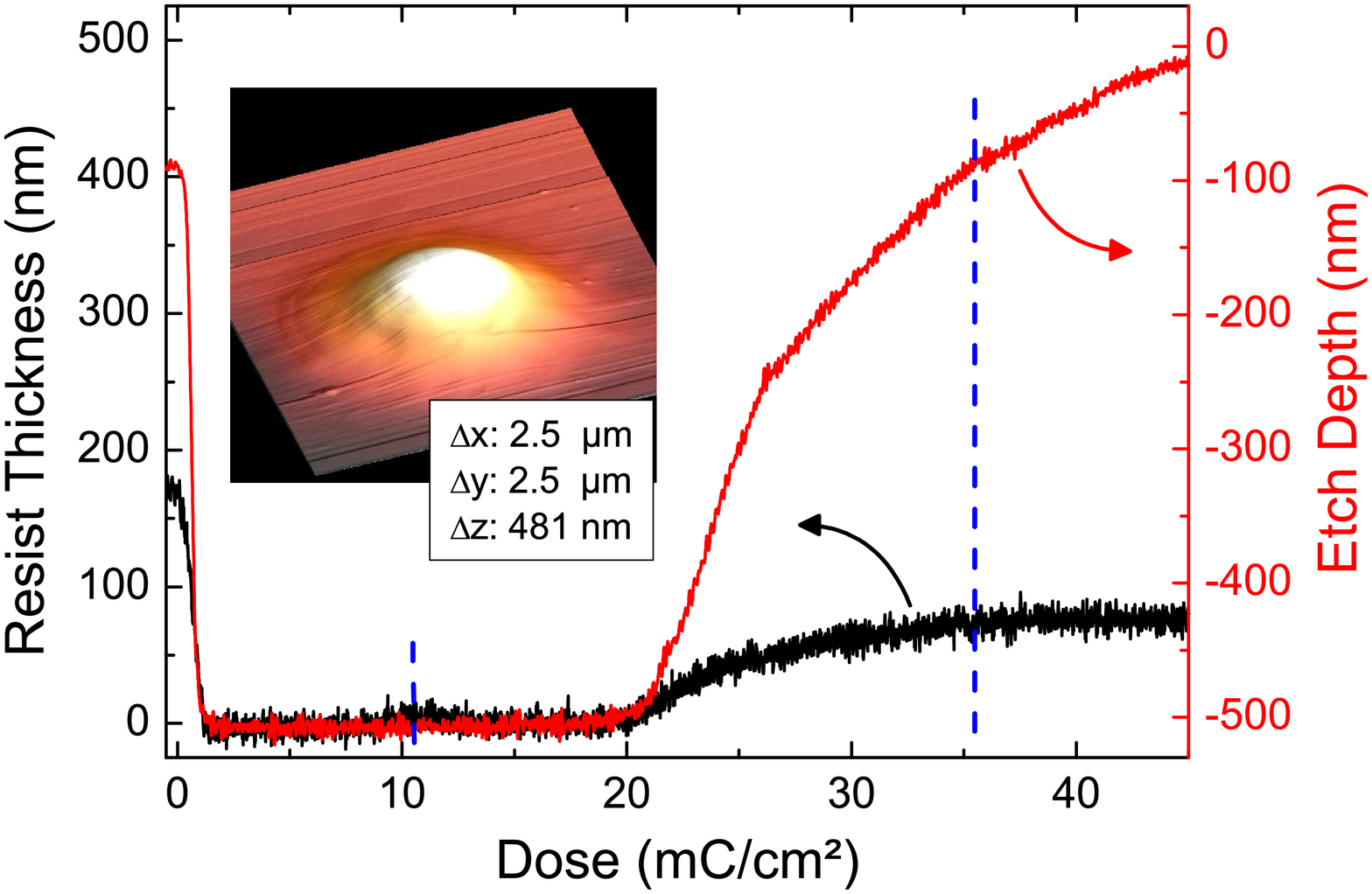}
\end{figure}
\begin{center}Gschrey et al. Fig. 2.\end{center}

\clearpage

\begin{figure}[h]
\centering\includegraphics[width=8 cm]{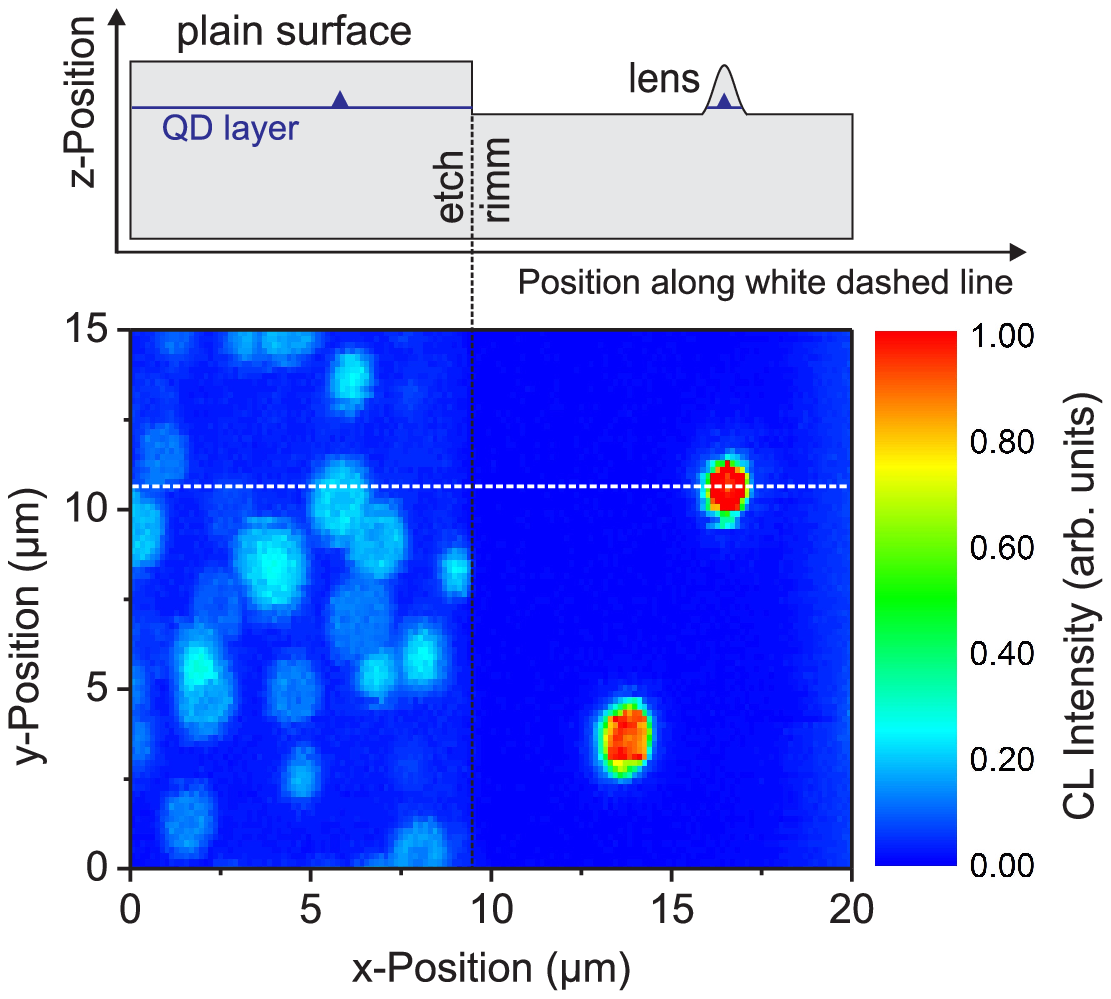}
\end{figure}
\begin{center}Gschrey et al. Fig. 3.\end{center}

\clearpage

\begin{figure}[h]
\centering\includegraphics[width=8 cm]{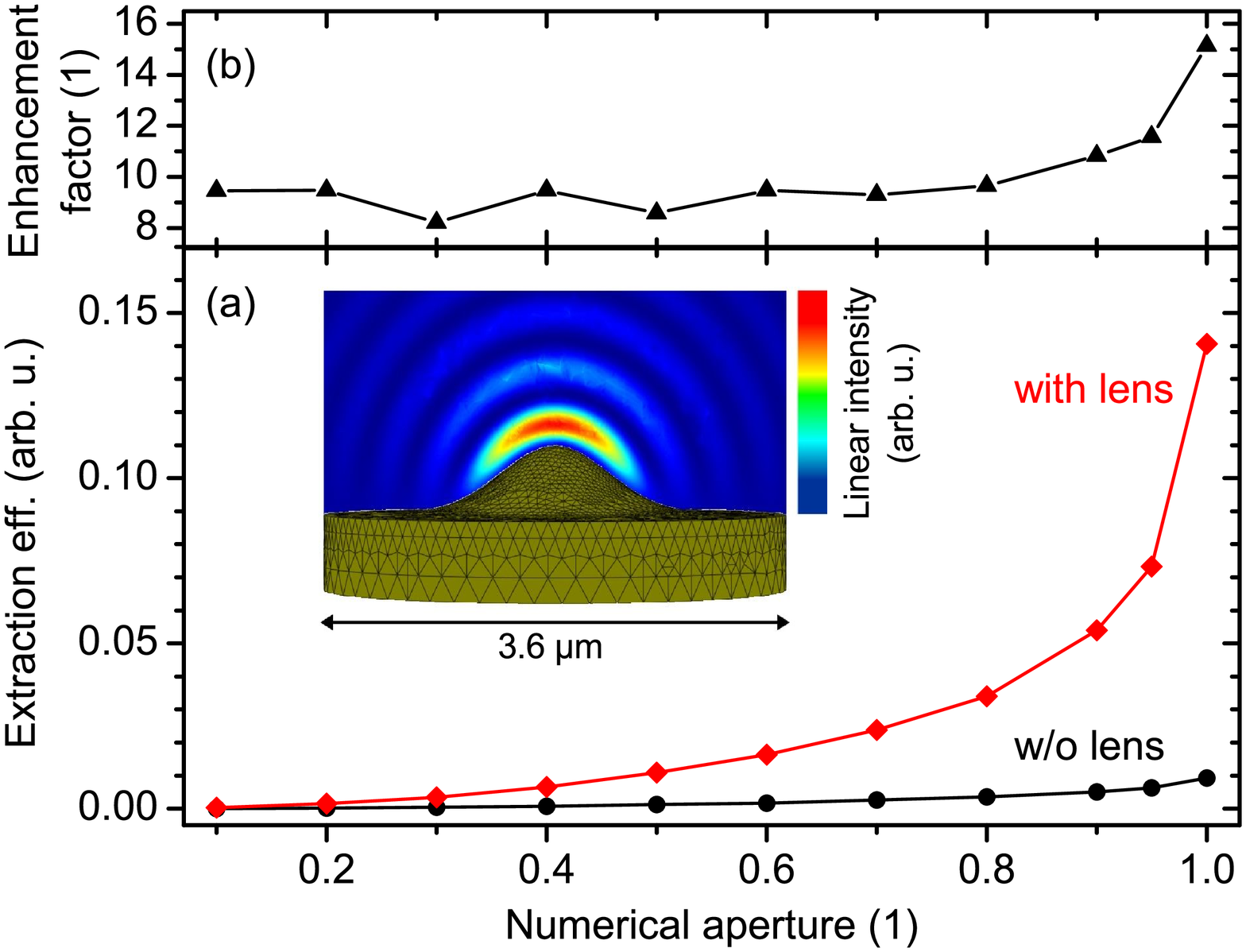}
\end{figure}
\begin{center}Gschrey et al. Fig. 4.\end{center}

\clearpage

\begin{figure}[h]
\centering\includegraphics[width=8 cm]{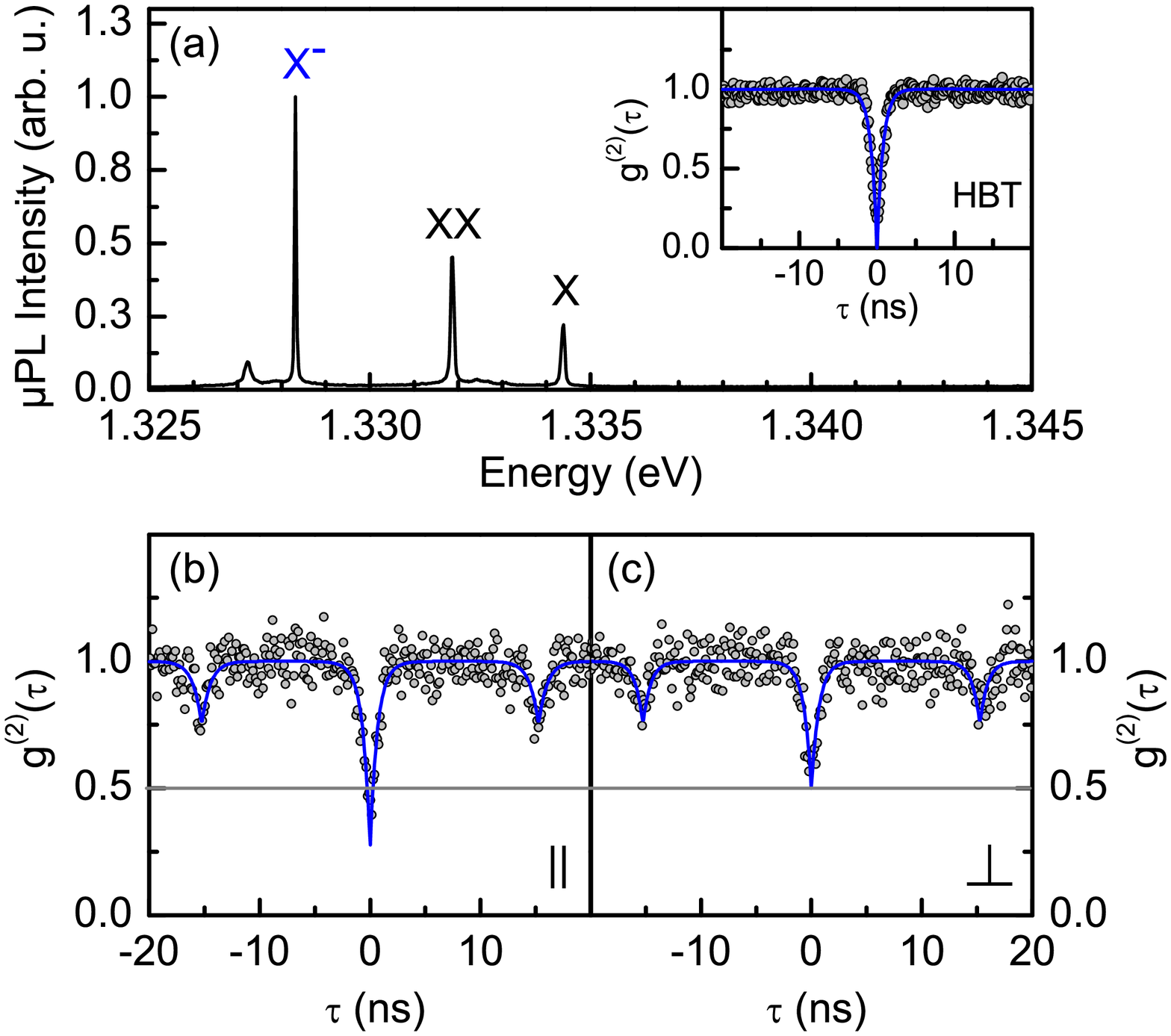}
\end{figure}
\begin{center}Gschrey et al. Fig. 5.\end{center}

% \end{document}

\clearpage

\section{Supplementary Information}

\subsection{Sample growth and structure}

The sample was grown by metal-organic chemical vapor deposition on GaAs(001) substrate. First, 500 nm of GaAs were deposited followed by 20~nm of Al$_{0.6}$Ga$_{0.4}$As as diffusion barrier for charge carriers and 150~nm of GaAs. The self-organized QDs were formed at a temperature of 500$\,^{\circ}\mathrm{C}$ during a growth interruption of 35~s. Finally the QDs were capped by 150~nm of GaAs, 20~nm of Al$_{0.4}$Ga$_{0.6}$As as second diffusion barrier and 230 nm GaAs as top layer. The rather thick top layer provides the material for the final lens structures after etching. 
%Due to intrinsic carbon doping during growth a p-type background doping of $\approx 1 \cdot 10^{17}$~cm$^{-3}$ is present in the sample.

\subsection{Determination of enhancement factor}

To determine the enhancement factor we performed a statistical analysis in which we compare the CL emission intensity of a large number (25) of QDs in the unpatterned region with the QDs below our microlenses. These measurements  were conducted as a function of excitation density. The reference intensity for each QD was the intensity of the single-excitonic lines when they saturated. This procedure was chosen to rule out the effect of different excitation conditions for QDs in microlenses and in the unprocessed regions. The analysis yields an enhancement factor as high as 6.6 $\pm$ 2.7 for ideally shaped microlenses for NA = 0.8.

\subsection{Hanbury-Brown and Twiss, and Hong-Ou-Mandel experiments}

For details of the HBT setup please refer to~\cite{Gsc13}.The Hong-Ou-Mandel experiment was carried out using an asymmetric Mach-Zehnder interferometer based on polarization maintaining fibers and a half wave plate (HWP) for switching the polarization in one arm of the interferometer~\cite{michler2009, ates2012two}. Silicon avalanche photo diodes with a timing resolution of 350~ps acted as detectors. In the measurements also two side dips at $\pm15.25$~ns down to 0.75 are observed. These dips originate from the optical delay in one interferometer arm and the symmetry of the side dips indicate a balanced second beam splitter ($R_{2}=T_{2}=50 \%$ )~\cite{Ates2009a}. All measurements were performed at a temperature of 15~K with an excitation power of 
$P=2.1$~$\mu$~W corresponding to low power-power excitation well below saturation of the QD states.

\section{Numerical method}

The calculation of the photon-extraction efficiency and of the enhancement factor was performed in the framework of a finite-element method by using the commercially available software package {\it JCMsuite} by JCMwave (see http://www.jcmwave.com for details). Based on AFM results, a full 3D model structure of the lens was created and the light intensity distribution resulting from point sources placed in the semiconductor material below the microlens was computed. The angular integration of far field intensity for a given NA was performed in a post-processing step.

\end{document}